# Milky Way Mass Through Escape Velocity Curve from LAMOST K Giants

Yin Wu[1,2], Haining Li[1], Yang Huang[2], Xiang-Xiang Xue[1], and Gang Zhao[1,2]
[1] CAS Key Laboratory of Optical Astronomy, National Astronomical Observatories, Chinese Academy of Sciences, Beijing 100101, People's Republic of China; lhn@nao.cas.cn
[2] School of Astronomy and Space Science, University of Chinese Academy of Sciences, No.19(A) Yuquan Road, Shijingshan District, Beijing, 100049, People's Republic of China

## Abstract

Escape velocity has long been used to constrain the mass of the dark matter (DM) halo in the Milky Way (MW). Here, we present a study of the escape velocity curve using a sample of high-velocity K giants with full 6D phase-space information and of relatively good quality, selected from LAMOST DR8 and cross-matched with Gaia DR3. To expand the high-velocity stars to larger distances, we used radius-dependent criteria of total velocity, i.e., $v_{\rm GC} > 300 \ {\rm km \ s^{-1}}$ for the solar neighborhood and $v_{\rm GC} > v_{\rm min} \sim 0.6 \times v_{\rm esc}(r_{\rm GC})$ for the outer region. We also selected halo stars based on $v_\phi - {\rm [Fe/H]}$ information to ensure that the sample is isotropic. We modeled the velocity distribution with traditional power-law models to determine the escape velocity in each radial bin. For the first time, we have directly measured a relatively continuous escape velocity curve that can extend to Galactocentric radii of ∼50 kpc, finding a decline in agreement with previous studies. The escape velocity at the solar position yielded by our measurements is $523.74^{+12.83}_{-13.47}$ km s$^{-1}$. Combined with the local circular velocity, we estimated the mass of the MW assuming a Navarro–Frenk–White DM profile, which resulted in a total mass of $M_{200, {\rm total}} = 0.90^{+0.06}_{-0.07} \times 10^{12} \ M_\odot$, with a concentration of $c_{200} = 13.47^{+1.85}_{-1.70}$. The small uncertainty implies that including the escape velocities beyond the solar neighborhood can result in a more precise mass estimate. Our derived MW mass is consistent with some recent studies using the escape velocity as well as other tracers, which may support a lower mass of the DM halo than in the past.

*Unified Astronomy Thesaurus concepts:* Milky Way dynamics (1051); Stellar kinematics (1608); Galaxy stellar halos (598)

## 1. Introduction

In the Milky Way (MW), dark matter (DM) dominates the mass of our galaxy (S. M. Faber & J. S. Gallagher 1979) and is expected to exceed the contribution of visible baryonic (stars and gas) components by more than an order of magnitude (J. Einasto et al. 1974; J. P. Ostriker et al. 1974). Since the existence of DM has been confirmed, the total mass and density profile of the DM halo have become crucial when modeling the formation and evolution of the MW (e.g., J. Bland-Hawthorn & O. Gerhard 2016), which can provide important observational constraints for the particle nature (e.g., S. Tulin & H.-B. Yu 2018; E. O. Nadler et al. 2021) and the indirect detection of DM (e.g., M. Cirelli et al. 2011; H. Abdallah et al. 2016). However, the mass of the DM halo has not been well constrained so far, with an estimated total mass around $10^{12} M_\odot$ and an uncertainty of at least a factor of 2 (see, e.g., W. Wang et al. 2020; S. A. Bird et al. 2022; T. Sawala et al. 2023, for recent reviews).

Numerous methods have been applied to measure the mass of the MW, such as the circular velocity curve (e.g., Y. Huang et al. 2016; A.-C. Eilers et al. 2019; Y. Zhou et al. 2023; X. Ou et al. 2024; F. Sylos Labini 2024), the population and motion of satellites (e.g., E. Patel et al. 2018; M. K. Rodriguez Wimberly et al. 2022), stellar streams (S. L. J. Gibbons et al. 2014; K. Malhan et al. 2018; E. Vasiliev et al. 2021), and the escape velocity of the MW (e.g., M. C. Smith et al. 2007; T. Piffl et al. 2014; A. A. Williams et al. 2017; A. J. Deason et al. 2019). All of them are based on the dynamics of halo stars, globular clusters, and dwarf satellite galaxies. Among these methods, the escape velocity, i.e., the maximum velocity for a star to be bound to the MW, can better constrain the mass beyond the measured radius because it retains gravitational potential information at both the measured radius and distant positions.

The escape velocity method has long been used to measure the mass of the MW halo since P. J. T. Leonard & S. Tremaine (1990) proposed a standard approach, which obtains the escape velocity by modeling the high-velocity tail of the stellar velocity distribution as a power-law model. Due to a lack of proper motion measurements, some early works only used line-of-sight velocity to select high-velocity halo stars suitable for escape velocity measurement. For example, M. C. Smith et al. (2007) and T. Piffl et al. (2014) estimated the escape velocity in the solar neighborhood, using the line-of-sight velocity from the Radial Velocity Experiment (RAVE; M. Steinmetz et al. 2006), finding that the local escape velocity lies in the range of [500–600] km s$^{-1}$. Combined with the mass model of the MW, these two works reported larger mass estimates in the range of (1–2) × $10^{12} M_\odot$. A. A. Williams et al. (2017) applied a similar approach to high-velocity samples from the Sloan Digital Sky Survey (SDSS; D. G. York et al. 2000), which first provided the measurement of escape velocity over a wide radius range extending out to Galactocentric radii of ∼50 kpc. Their results are consistent with the previous local escape velocity measurements and show that the escape velocity decreases to 379 km s$^{-1}$ at a radius of 50 kpc.

The Gaia satellite provides unprecedented accurate astrometry, including positions, parallax, proper motions, and radial velocities (Gaia Collaboration et al. 2016). Combined with other large spectroscopic surveys, such as RAVE (M. Steinmetz et al. 2006), SDSS (D. G. York et al. 2000), LAMOST (X.-Q. Cui et al. 2012; G. Zhao et al. 2012), APOGEE (S. R. Majewski et al. 2017), GALAH (G. M. De Silva et al. 2015), etc., it is possible to select high-velocity stars with full 6D phase-space information through the total velocity (e.g., H. H. Koppelman & A. Helmi 2021; L. Necib & T. Lin 2022a; Z. Prudil et al. 2022; C. Roche et al. 2024), which provides a new opportunity to measure the escape velocity, and subsequently the mass and distribution of the DM. The number and measured radius of high-velocity star samples have been greatly expanded, allowing more precise determinations of the escape velocity within a few kiloparsecs and the view of an overall escape velocity curve (e.g., G. Monari et al. 2018; H. H. Koppelman & A. Helmi 2021; C. Roche et al. 2024).

In the post-Gaia era, there have been many studies focusing on the escape velocity in the solar neighborhood (G. Monari et al. 2018; A. J. Deason et al. 2019; R. J. J. Grand et al. 2019; H. H. Koppelman & A. Helmi 2021; L. Necib & T. Lin 2022a; C. Roche et al. 2024). The estimate of the local escape velocity with Gaia DR2 is still in [500–640] km s$^{-1}$ (G. Monari et al. 2018; A. J. Deason et al. 2019), while for some studies with Gaia DR3 data, it tends to be smaller than before ($\sim$480 km s$^{-1}$, H. H. Koppelman & A. Helmi 2021; L. Necib & T. Lin 2022a; C. Roche et al. 2024). Subsequently, the mass of the DM halo tends downward according to the recent results. Differences in these studies, such as the treatment of the model parameters (e.g., L. Necib & T. Lin 2022a, 2022b; C. Roche et al. 2024) and prior assumptions (e.g., M. C. Smith et al. 2007; T. Piffl et al. 2014; A. J. Deason et al. 2019), may play a role in the resulting escape velocity. One of the most dominant reasons is the different statistics. Gaia DR3 includes more sources at large distances with radial velocities than Gaia DR2, as well as the pre-Gaia studies. Therefore, using a new data set to measure the escape velocity is necessary to further understand the detailed distribution of the DM.

Furthermore, few studies have involved the measurement of escape velocities in the outer region. For example, A. A. Williams et al. (2017) used various tracers, including main-sequence turn-off stars, blue horizontal branch stars, and K giants, to model the high-velocity tail over a wide radius range. Their work did not directly measure the escape velocity within small radial bins but added the radial variation of the escape velocity into the simulation. Recently, using RR Lyrae variables, Z. Prudil et al. (2022) measured the escape velocity that extended to Galactocentric radii of 28 kpc. Due to the limited number of distant stars, their work yielded results only for three radial bins. Thus, a relatively continuous escape velocity curve extending to outer region is required. Such a curve can provide an overall view of the DM mass distribution, and thereby help us better understand the evolution and structure of the MW.

In this work, we attempt to obtain an escape velocity curve that can extend to the outer region and estimate the DM halo mass utilizing another stellar tracer, K giants selected from LAMOST (X.-Q. Cui et al. 2012; G. Zhao et al. 2012) and cross-matched with Gaia DR3 (Gaia Collaboration et al. 2023). The high-velocity subset in our K-giant sample extends out to Galactocentric radii of $\sim$50 kpc, which makes it appropriate to investigate the escape velocity in the outer region. The main idea is to use the methodology described by P. J. T. Leonard & S. Tremaine (1990) for the escape velocity at a given radius, but to add the contribution of outliers following recent studies (e.g., A. A. Williams et al. 2017; L. Necib & T. Lin 2022a; C. Roche et al. 2024). With the measured escape velocity curve, the virial mass of the MW can be fitted with a mass distribution of the MW. This paper is organized as follows. In Section 2, we introduce the K-giant sample and the selection of high-velocity stars. The method for estimating the escape velocity is described in Section 3. The results are summarized in Section 4 and the mass estimate of the DM halo is presented in Section 5. Finally, we present our conclusions in Section 6.

## 2. Data

In order to obtain the escape velocity curve extending to outer region, we used the K giants selected from LAMOST Data Release 8 (DR8). LAMOST has obtained numerous low-resolution spectra covering a wavelength range of $3700 < \lambda < 9000$ Å and provided accurate metallicity and line-of-sight velocity measurements (X.-Q. Cui et al. 2012; G. Zhao et al. 2012). Due to the fact that K giants are quite bright, a large sample of K giants has been observed in LAMOST DR8 (e.g., L. Zhang et al. 2023). These K giants were identified by the selection criteria based on the effective temperature $T_{\rm eff}$ and surface gravity $\log g$ (C. Liu et al. 2014), i.e., $\log g < 3.5$ when $4000 < T_{\rm eff}/{\rm K} < 4600$ or $\log g < 4$ when $4600 \leqslant T_{\rm eff}/{\rm K} < 5600$. The distance ($d$) was estimated with the Bayesian method described in X.-X. Xue et al. (2014), which provides unbiased distance estimates for distant halo stars. We cross-matched the K-giant sample with the Gaia DR3 catalogs (Gaia Collaboration et al. 2023) for accurate proper motion measurements. The cross-matched sample contains 647,441 stars, among which 614,294 stars have full 6D information.

We then calculated the space velocities and associated uncertainties for all stars with full 6D information, adopting the radial velocities from LAMOST DR8 and the proper motions from Gaia DR3 and using the data processing pipeline of C. Roche & L. Necib (2023). The distance from the solar position to the Galactic center is $R_\odot = 8.122$ kpc (GRAVITY Collaboration et al. 2018), and the position above the Galactic plane is $z_\odot = 20.8$ pc (M. Bennett & J. Bovy 2019). The velocities were corrected for the solar peculiar motion and the local circular velocity, which are $(U_\odot, V_\odot, W_\odot) = (11.1, 12.24, 7.25)$ km s$^{-1}$ (R. Schönrich et al. 2010) and $v_{\rm circ}(R_\odot) = 234.04$ km s$^{-1}$ (Y. Zhou et al. 2023). We used a right-handed Galactocentric Cartesian coordinate system $(X, Y, Z)$, with $X$ toward the Galactic Center from the Sun, $Y$ toward the direction of the MW's rotation, and $Z$ pointing to the North Galactic Pole. We also adopted a right-handed Galactocentric cylindrical system $(R, \phi, Z)$, indicating negative $v_\phi$ for the stars with prograde orbits. The total velocity and Galactocentric radius are $v_{\rm GC} = \sqrt{v_R^2 + v_\phi^2 + v_Z^2}$ and $r_{\rm GC} = \sqrt{X^2 + Y^2 + Z^2}$, respectively. Since the determination of escape velocity is highly dependent on the velocity measurement, we selected relatively good quality stars through the criteria of $\sigma_d/d < 50\%$ and $\sigma_{v_{\rm GC}} < 50$ km s$^{-1}$. There are 586,291 stars ($\sim$90% of the original sample) available for further selection of high-velocity stars.

With the total velocity calculated above, we proceed to select high-velocity stars. In previous studies, the selection of high-velocity stars in the solar neighborhood has always taken a fixed lowest threshold ($v_{\rm min}$) between 250 km s$^{-1}$ and 300 km s$^{-1}$ (e.g., G. Monari et al. 2018; A. J. Deason et al. 2019). However, such criteria may not be suitable for selecting high-velocity stars outside the solar neighborhood. The upper boundary of the high-velocity tail depends on the escape velocity, which decreases with increasing Galactocentric radius. When using a fixed $v_{\min}$, it is impossible to select adequate high-velocity stars in the outer region for escape velocity measurement. Therefore, we chose to use radius-dependent selection criteria based on the theoretical escape velocity curve, similar to J. Liao et al. (2024).

First, we explored the modeled Galactic escape velocity curve using the MW potential implemented in `galpy`[3] (MWPotential2014, J. Bovy 2015). From the `galpy` module, the escape velocity is defined as:

$$v_{\rm esc}(r_{\rm GC}) = \sqrt{2|\Phi(r_{\rm GC}) - \Phi(\infty)|}, \quad (1)$$

where the infinity is described as the radius at $10^{12}$ with an arbitrary unit of $1.0/R_\odot$. The MWPotential2014 includes a bulge modeled as a power-law density profile with an exponential cutoff, a Miyamoto–Nagai stellar disk (M. Miyamoto & R. Nagai 1975), and a DM halo modeled as a Navarro–Frenk–White (NFW) profile (J. F. Navarro et al. 1996, 1997). A detailed description of the parameters and physical properties can be found in Table 1 of J. Bovy (2015). Based on recent research results, we set the solar distance to the Galactic center to $R_\odot = 8.122$ kpc (GRAVITY Collaboration et al. 2018) and the solar circular velocity to $v_{\rm circ}(R_\odot) = 234.04$ km s$^{-1}$ (Y. Zhou et al. 2023). To consider the effect of DM mass, we derived escape velocity curves with $M_{\rm vir} = 0.5 \times 10^{12} M_\odot$ (MWPotential2014$^{\rm L}$), $M_{\rm vir} = 0.85 \times 10^{12} M_\odot$ (MWPotential2014$^{\rm M}$), and $M_{\rm vir} = 1.2 \times 10^{12} M_\odot$ (MWPotential2014$^{\rm H}$), covering the recent mass range calculated by different methods (see Figure 8 in S. A. Bird et al. 2022).

Second, we selected high-velocity stars based on $v_{\rm GC}$. For the solar neighborhood ($4 \leqslant r_{\rm GC} \leqslant 12$ kpc), high-velocity stars are defined as $v_{\rm GC} > 300$ km s$^{-1}$, in line with the literature (e.g., A. J. Deason et al. 2019; C. Roche et al. 2024). For the outer region ($r_{\rm GC} > 12$ kpc), high-velocity stars are defined as $v_{\rm GC} > v_{\min} \sim 0.6 \times v_{\rm esc}(r_{\rm GC})$, where $v_{\min}$ is the low bound. The scale factor 0.6 comes from the extrapolation of 300 km s$^{-1}$ with respect to the theoretical escape velocity curve. Furthermore, we excluded extremely fast stars with $v_{\rm GC} > 700$ km s$^{-1}$ because these extreme stars cannot be bound, even taking into account the measurement errors.

Finally, we obtained a subset with ∼5000 high-velocity K giants for the escape velocity measurement. These selected stars have relatively good quality, with a typical uncertainty of 17% in $d$ and 8% in $v_{\rm GC}$. To test the consistency of the escape velocity curve, we divided these stars into two different bins, Bin 1 and Bin 2, with the same bin width but an offset of 0.5 kpc.

Figure 1 shows the $v_{\rm GC} - r_{\rm GC}$ and $v_\phi$ - [Fe/H] diagrams of the selected stars, which are divided into Bin 1 and Bin 2. As shown in the $v_{\rm GC} - r_{\rm GC}$ diagrams (top panels of Figure 1), the selected high-velocity K-giant sample has only a small number of unbound stars, which is suitable for the P. J. T. Leonard & S. Tremaine (1990) method. The radius-dependent selection results in a larger high-velocity star sample that can extend to the outer radius. In our study, even in the outermost radial bin ($r_{\rm GC} \in [40, 48]$ kpc), there are ∼50 stars for the escape velocity measurement. We note that two clear clumps are clearly shown in the $v_\phi$ - [Fe/H] diagrams: the halo population with low metallicity clustered at $v_\phi \sim 0$ km s$^{-1}$, and the disk population with high metallicity clustered at $v_\phi \sim -300$ km s$^{-1}$. To further eliminate disk contamination, we applied an empirical criteria, $v_\phi = 276.70 \times$ [Fe/H] + 97.78, the same as Q.-Z. Li et al. (2023), where 2977 (Bin 1) and 2891 (Bin 2) stars are classified as halo stars. All the cuts and the number of stars left are summarized in Table 1.

---
[3] Available at http://github.com/jobovy/galpy.

## 3. Methods

### 3.1. Modeling Approach

To determine the escape velocity ($v_{\rm esc}$), we used the methodology first proposed by P. J. T. Leonard & S. Tremaine (1990), which assumes that the velocity distribution of stars can be described by a power-law model with the following form:

$$p(v_{\rm GC}|v_{\rm esc}, k) \propto (v_{\rm esc} - v_{\rm GC})^k \quad (2)$$

for $v_{\min} \leqslant v_{\rm GC} < v_{\rm esc}$. Here, $v_{\rm GC}$ is the total Galactocentric velocity, $k$ is the exponent, $v_{\rm esc}$ is the escape velocity, and $v_{\min}$ is the cutoff (usually between 250 km s$^{-1}$ and 300 km s$^{-1}$ in the solar neighborhood) beyond which the $v_{\rm GC}$ distribution can still be described by Equation (2).

When adopting this methodology, the stellar velocity distribution implicitly implies the following assumptions. First, the stellar velocity distribution is assumed to be a smooth function extending to $v_{\rm esc}$, i.e., the velocity distribution does not truncate at a low value. Second, the velocity distribution would truncate at $v_{\rm esc}$, thus there are no unbound stars. Furthermore, since the velocity distribution only depends on velocity, implying an isotropic distribution, we assume that in an appropriate sample there are no rotating (disk-like) stars that would break the isotropy on the sky.

The first assumption is often broken because the distribution is not always smooth due to substructures and is often truncated below the true $v_{\rm esc}$. For example, M. C. Smith et al. (2007) proposed that, based on cosmological simulations, velocity distributions can reach 90% of $v_{\rm esc}$. Recent results also show that the above methodology tends to underestimate $v_{\rm esc}$ by 7%–10% (R. J. J. Grand et al. 2019; H. H. Koppelman & A. Helmi 2021).

The latter two assumptions are easily ensured for our sample. As demonstrated in Section 2, we adopted the criteria of metallicity and $v_\phi$ to select halo stars, which ensured an isotropic distribution. Since unbound stars are typically young stars ejected from the Galactic center and not old stars in the halo (e.g., W. R. Brown 2015), our sample is unlikely to contain many unbound stars, as shown in Figure 1. Nevertheless, following previous studies (e.g., A. A. Williams et al. 2017; L. Necib & T. Lin 2022a, 2022b), we include a wide Gaussian outlier model to describe possible unbound stars or mismeasured stars:

$$p_{\rm out}(v_{\rm GC}) \propto \exp\left(-\frac{v_{\rm GC}^2}{2[\sigma_{v_{\rm GC}}^2 + \sigma_{\rm out}^2]}\right), \quad (3)$$

where $\sigma_{v_{\rm GC}}$ is the total velocity error calculated in Section 2 and $\sigma_{\rm out}$ is fixed to 1000 km s$^{-1}$. After introducing a free factor $f$ for the fraction of outliers, the total likelihood function is given

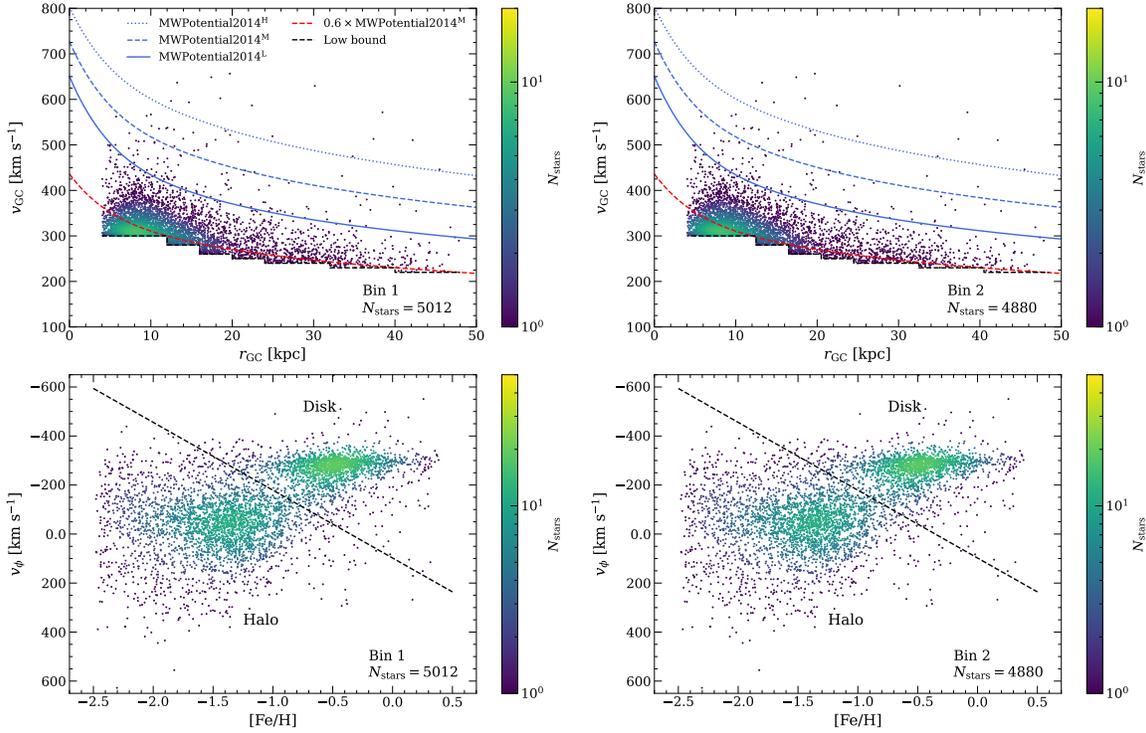

**Figure 1.** Top panels: The density distribution of high-velocity K giants divided into Bin 1 (left panel) and Bin 2 (right panel) on the plane of $v_{\rm GC}$ and $r_{\rm GC}$. The blue lines represent the escape velocity curve calculated using a gravitational potential for the MW (MWPotential2014, J. Bovy 2015) with different virial masses. The red dashed line represents 60% of the average escape velocity curve. The black dashed line marks the low bound of $v_{\rm GC}$ for high-velocity stars. Bottom panels: The distribution of high-velocity K giants on the [Fe/H] − $v_\phi$ plane for Bin 1 (left panel) and Bin 2 (right panel), respectively. The black dashed line represents the criteria we use to select the halo population (lower left).

**Table 1**
Summary of Cuts Applied to Select High-velocity Halo Stars with Good Quality

| Cuts | $N_{\rm stars}$ |
| --- | --- |
| Original K giants from LAMOST DR8 | 652,027 |
| Cross-matched with Gaia DR3 | 647,441 |
| Full 6D phase-space information | 614,294 |
| $\sigma_d/d < 50\%$ | 597,749 |
| $\sigma_{v_{\rm GC}} < 50$ km s$^{-1}$ | 586,291 |
| High-velocity stars, $r_{\rm GC} \in [4, 48]$ kpc | 5017 (Bin 1)/4885 (Bin 2) |
| $v_{\rm GC} < 700$ km s$^{-1}$ | 5012 (Bin 1)/4880 (Bin 2) |
| Halo stars | 2977 (Bin 1)/2891 (Bin 2) |

by:

$$P_{\rm total} = \prod_i^{N_{\rm stars}} [(1-f)p_i(v_{\rm GC}|v_{\rm esc}, k) + f p_{{\rm out},i}(v_{\rm GC})]. \quad (4)$$

During the determination of escape velocity, we assume that $v_{\rm esc}$ is approximately constant over a sufficiently small radius range. Since we do not know the specific range in which this assumption holds, it is necessary to introduce the variation of $v_{\rm esc}$ in each radial bin. We take into account changes in $v_{\rm esc}$ over radii following the prescription by A. A. Williams et al. (2017) and A. J. Deason et al. (2019):

$$v_{\rm esc}(r_{\rm GC}) = v_{\rm esc}(r_0)(r_{\rm GC}/r_0)^{-\gamma/2}, \quad (5)$$

where $r$ is the Galactocentric radius, $r_0$ is the median value of the observed Galactocentric radius for the stars in each radial bin, and $v_{\rm esc}(r_0)$ is the escape velocity at $r_0$. The parameter $\gamma$ is free and represents the slope of the gravitational potential.

Finally, the Bayesian theorem can be applied to the likelihood function and the probability of the model parameters:

$$\begin{aligned} P(v_{\rm esc}, k, \gamma, f | v_{\rm GC}) &\propto \\ P(v_{\rm esc})&P(k)P(\gamma)P(f)P_{\rm total}(v_{\rm GC}|v_{\rm esc}(r_0), k, \gamma, f), \end{aligned} \quad (6)$$

where $P(v_{\rm esc})$, $P(k)$, $P(\gamma)$, and $P(f)$ represent the priors on $v_{\rm esc}(r_0)$, $k$, $\gamma$, and $f$, respectively.

### 3.2. Detailed Settings

We then describe the detailed settings in our modeling process, such as the priors, initial values, and numerical implementation. The priors and initial values for the modeling $v_{\rm esc}$, $k$, $\gamma$, and $f$ are given in Table 2. Following some previous studies (e.g., T. Piffl et al. 2014; H. H. Koppelman & A. Helmi 2021), we adopted a simple prior on $v_{\rm esc}$ of the form $P(v_{\rm esc}) \propto 1/v_{\rm esc}$. For other parameters, we assumed a uniform prior function as in L. Necib & T. Lin (2022a, 2022b) and C. Roche et al. (2024).

The exponent $k$ is predicted by the distribution function of the energy. For a given distribution function, it is possible to obtain an exact value of $k$, such as $k = 3.5$ for a Plummer model (H. C. Plummer 1911). However, for a real galaxy such as the MW, it is difficult to constrain a precise $k$ value. The prior assumption on the range of $k$ significantly affects the resulting posterior distribution (M. C. Smith et al. 2007; G. Monari et al. 2018; A. J. Deason et al. 2019). The larger $k$ corresponds to a flat velocity distribution and may bias larger $v_{\rm esc}$ values. A large sample ($N_{\rm stars} > 200$) with very small uncertainties is necessary to estimate $v_{\rm esc}$ and $k$ simultaneously (P. J. T. Leonard & S. Tremaine 1990; T. Piffl et al. 2014).

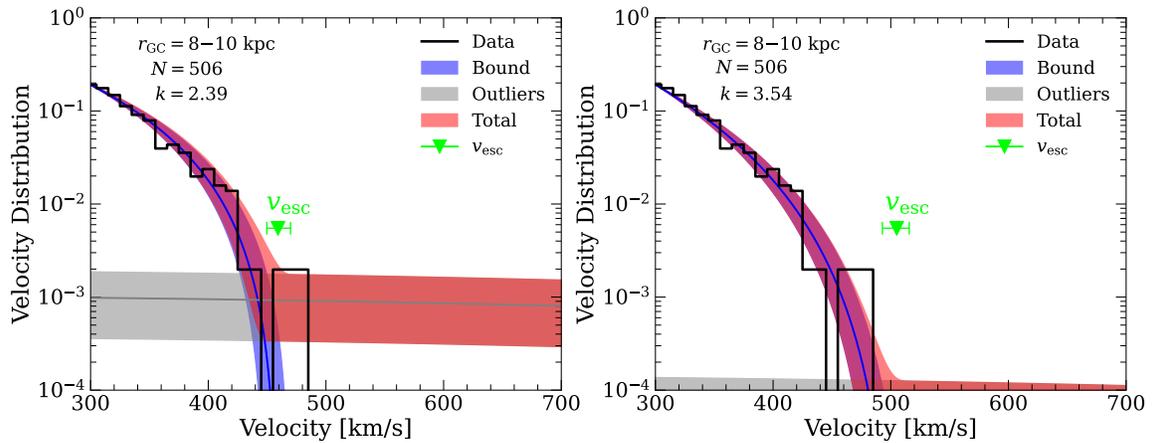

**Figure 2.** Comparison of best-fit results to velocity distribution of stars using the prior of $k \in [0.1, 2.5]$ (left panel) and $k \in [0.1, 3.7]$ (right panel) at $v_{\min} = 300$ km s$^{-1}$ for a representative radial bin at 8–10 kpc. The black solid line is the velocity distribution of stars, while the shaded regions are the distributions predicted by the model with 68% confidence intervals. The contributions of the bound and outlier components are shown individually. The $v_{\rm esc}$ marker with error bars represents the best-fit escape velocity and its 68% confidence intervals.

**Table 2**
Priors and Initial Values for Modeling

| Parameter | Prior Range | Prior | Initial Value |
|---|---|---|---|
| $v_{\rm esc}(r_0)$ | $[v_{\min}, 1000]$ km s$^{-1}$ | $1/v_{\rm esc}(r_0)$ | 500 km s$^{-1}$ |
| $k$ | [0.1, 2.5] or [0.1, 3.7] | 1 | 2.0 or 3.0 |
| $\gamma$ | [0.0, 1.0] | 1 | 0.1 |
| $f$ | $[10^{-6}, 1.0]$ | 1 | $10^{-5}$ |

Therefore, due to the limited sample in the outer radial bins, we chose to adopt the prior range of $k$ from the literature.

A series of works constrained $k$ through cosmological simulations (P. J. T. Leonard & S. Tremaine 1990; M. C. Smith et al. 2007; T. Piffl et al. 2014; A. J. Deason et al. 2019). For example, in the seminal work of P. J. T. Leonard & S. Tremaine (1990), a lower range [0.5–2.5] was preferred to include the system with $k = 1.5$ that has undergone violent relaxation (L. A. Aguilar & S. D. M. White 1986; W. Jaffe 1987; S. Tremaine 1987). After that, M. C. Smith et al. (2007) assumed a larger range ($k \in [2.3, 4.7]$) based on cosmological simulations of MW-like galaxies, while T. Piffl et al. (2014) updated the range to [2.3–3.7]. Recently, A. J. Deason et al. (2019) argued that the range [1.0–2.5] is more favored using the Auriga suite regarding the merger history of the MW. Due to the difficulty of determining $k$, we chose to adopt two different $k$ prior ranges based on recent simulations, [0.1–2.5] and [0.1–3.7], to better model the shape of the velocity distribution.

For the two $k$ prior ranges, [0.1–2.5] and [0.1–3.7], we utilized the Markov Chain Monte Carlo (MCMC) sampler to find the posterior distribution of $k$ and $v_{\rm esc}$, respectively. The MCMC sampling was performed in emcee (D. Foreman-Mackey et al. 2013), using 50 walkers, 2000 steps as a burn-in, and additional 3000 steps for further sampling. After obtaining the posterior distribution, we compared the goodness of fit via the AIC (H. Akaike 1974), defined as:

$$\text{AIC} = 2s - 2\log(\mathcal{L}_{\max}), \quad (7)$$

where $s$ is the number of model parameters (4 for our study), $\log \mathcal{L}_{\max}$ is the maximum of the log-likelihood in MCMC modeling. We then calculated $\Delta\text{AIC} \equiv \text{AIC}_{3.7} - \text{AIC}_{2.5}$, preferring $k \in [0.1, 2.5]$ if $\Delta\text{AIC} > 0$ and $k \in [0.1, 3.7]$ if $\Delta\text{AIC} < 0$, respectively. Since $s$ is the same in the modeling for the two $k$ prior ranges, we actually selected the results with larger $\mathcal{L}_{\max}$, i.e., the better-fit results.

## 4. Results

### 4.1. Escape Velocity Modeling Results

To obtain the overall escape velocity curve, we applied the above modeling procedure for each radial bin of both Bin 1 and Bin 2. The measured $v_{\rm esc}$ and its 68% credible intervals are listed in Appendix A (Table A1). The preferred fits for all radial bins are available in Appendix B (Figure B1 for Bin 1; Figure B2 for Bin 2). As we can see, the measurements for Bin 1 and Bin 2 are almost the same within the error range, implying that different bin manners have little effect on our results. Therefore, we only conduct a detailed analysis of the results for Bin 1 in the following.

Figure 2 shows the data and the best-fit velocity distribution for the two $k$ prior ranges, taking the 8–10 kpc radial bin as an example. The fits for $k \in [0.1, 2.5]$ and $k \in [0.1, 3.7]$ are shown in the left and right panels, respectively. The corresponding corner plots are shown in Appendix B (Figure B3). As demonstrated in Figure 2, the exponent $k$ may mainly govern the behavior of low-speed stars, especially the decreasing trend of the velocity distribution with the increase of $v_{\rm GC}$. When comparing the left and right panels, different $k$ priors can cause a difference of $\sim 30$ km s$^{-1}$ (see also, G. Monari et al. 2018; A. J. Deason et al. 2019). We can see that the fits for $k \in [0.1, 3.7]$ are more preferred than those for $k \in [0.1, 2.5]$, since the fits for $k \in [0.1, 3.7]$ can better reproduce the decreasing trend of the velocity distribution and the contribution of the outliers. The comparison of AIC ($\Delta\text{AIC}_{8-10\,\rm kpc} = -12.72 < 0$) also shows that $k \in [0.1, 3.7]$ is more favored for this radial bin. The possible reason is that a few stars with $v_{\rm GC} \sim 480$ km s$^{-1}$ are not well fitted when adopting smaller $k$ priors ([0.1–2.5]). As a result, the parameter $f$, i.e., the fraction of outliers, is overestimated, leading to a lower $v_{\rm esc}$.

In Figure B3, we note that the posterior distribution of $k$ shows a clear degeneracy between $v_{\rm esc}$ and $k$ in the 8–10 kpc radial bin. This may be due to the difficulty of distinguishing whether the stars with $v_{GC}$ very close to $v_{esc}$ are unbound. A consequently ideal assumption is that the degeneracy can be broken if the distribution of bound stars is well separated from that of possible unbound stars, which requires larger statistics with more strict quality cuts (e.g., ~5% velocity errors, L. Necib & T. Lin 2022a). In our study, we did not adopt such strict cuts, but chose the appropriate speed error truncation to ensure statistical data at the outer region. Nevertheless, in some radial bins, the degeneracy seems to be broken. The corner plots of the representative radial bins (14–16 kpc and 20–24 kpc) for this situation are shown in Appendix B (Figure B4). For these radial bins, $v_{esc}$ and $k$ can be modeled simultaneously when adopting appropriate $k$ prior range ($k \in [0.1, 3.7]$).

The shape of the escape velocity should be considered when discussing the modeling results. In previous studies, there are two main manners to describe the shape of the escape velocity curve. For example, a larger sample of stars can be divided into quite small radial bins (e.g., G. Monari et al. 2018; H. H. Koppelman & A. Helmi 2021; C. Roche et al. 2024). Therefore, the escape velocity is considered to be constant in the small bins. Another approach is to include a parameterized escape velocity profile in the modeling, for which the modeling can be performed over a larger radius range (e.g., A. A. Williams et al. 2017; A. J. Deason et al. 2019; Z. Prudil et al. 2022). We used the first approach, but introduced the parameter $\gamma$ to test the deviation caused by the variation of $v_{esc}$ within the small bins. As demonstrated in Figure B3, there is no evidence for a strong variation of $v_{esc}$ in the small radial bins ($\gamma \sim 0$), i.e., $v_{esc}$ is almost constant within the small bins. Therefore, we can conclude that introducing $\gamma$ has little effect on the modeling results, and the $v_{esc}(r_0)$ is credible in our work.

### 4.2. Escape Velocity Curve

With the escape velocity measured above, we demonstrate the results in different radial bins. The escape velocity curve for Bin 1 is shown in Figure 3. The error bar at the outermost bins would increase as a result of the decreasing sample size. While the error bar is quite small in the solar neighborhood, such as $548.93^{+9.68}_{-11.09}$ km s$^{-1}$ in the 6–8 kpc and $505.06^{+10.67}_{-12.20}$ km s$^{-1}$ in the 8–10 kpc. We interpolated the measurements and obtained the local escape velocity $v_{esc}(r_\odot) = 523.74^{+12.83}_{-13.47}$ km s$^{-1}$, which shows agreement with previous estimates at that position, especially for recent measurements based on large samples (e.g., H. H. Koppelman & A. Helmi 2021; L. Necib & T. Lin 2022a; C. Roche et al. 2024).

In Figure 3, we overplot the escape velocity curve predicted by A. A. Williams et al. (2017) and the measurements of escape velocities outside the solar position using RR Lyrae variables (Z. Prudil et al. 2022). A. A. Williams et al. (2017) performed the modeling in a distinct manner from our work, in which the escape velocities at the solar neighborhood and the outer region were obtained by modeling over a wide range of $r_{GC}$ instead of small radial bins. Their research led to the result $v_{esc} = 521^{+46}_{-30}$ km s$^{-1}$ at the local position and $v_{esc} = 379^{+34}_{-28}$ km s$^{-1}$ at $R = 50$ kpc. In Z. Prudil et al. (2022), a similar modeling approach was applied to the velocity distribution of RR Lyrae stars to measure the escape velocity at three radial bins extending to 28 kpc. Their measurements were $v_{esc} = 512^{+94}_{-37}$ km s$^{-1}$, $v_{esc} = 436^{+44}_{-22}$ km s$^{-1}$ and $v_{esc} = 393^{+53}_{-26}$ km s$^{-1}$ at 8 kpc, 16 kpc, and 24 kpc, respectively.

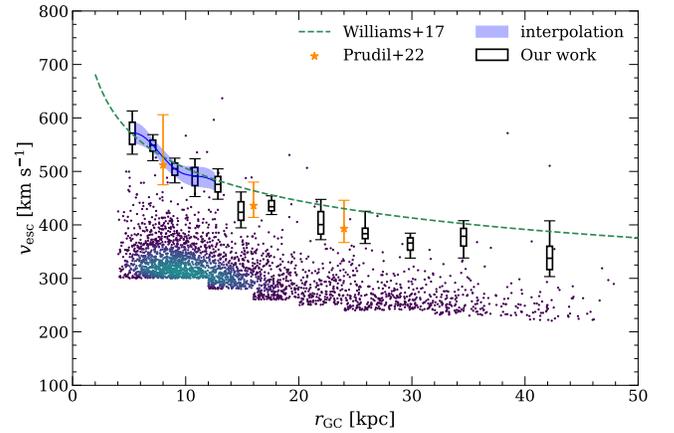

**Figure 3.** Comparison of the resulting escape velocity curves to previous escape velocity beyond the solar neighborhood. The escape velocity curves are shown as the escape velocity points with 68% (boxes) and 95% (whiskers) confidence intervals at different radii for Bin 1 (black). The orange points with error bars are the results of Z. Prudil et al. (2022). The green dashed line represents the escape velocity curve predicted by A. A. Williams et al. (2017). The interpolation results at the solar neighborhood also shows with 68% confidence intervals (blue).

Our results are well consistent with these measurements, showing a sharp decline of escape velocity at the inner region ($r_{GC} \lesssim 20$ kpc), while there is a relatively smooth decline at the outer region ($r_{GC} \gtrsim 20$ kpc).

## 5. Discussions

### 5.1. From Escape Velocity to Milky Way Potential

The escape velocity curve is related to the gravitational potential of a galaxy and can be defined as the velocity necessary for a star to escape to infinity, thus, $v_{esc}(r_{GC}) = \sqrt{2|\Phi(r_{GC})|}$. However, for a real galaxy such as the MW, we should choose a finite limiting radius at which a star is considered unbound. To remain consistent with the literature (A. J. Deason et al. 2019; L. Necib & T. Lin 2022a; C. Roche et al. 2024), the limiting radius is set to $2r_{200}$, where $r_{200}$ is the radius within which the average halo density is $200 \times \rho_{crit}$ (the critical density of the universe at redshift 0, equal to $3H^2/8\pi G$). In this work, we take the Hubble constant of $H = 70$ km s$^{-1}$ Mpc$^{-1}$ from recent measurements (W. L. Freedman et al. 2025). The above assumption leads to a modification in Equation (1) as:

$$v_{esc}(r_{GC}) = \sqrt{2|\Phi(r_{GC}) - \Phi(2r_{200})|}. \tag{8}$$

To estimate the mass of the MW, we need to model the escape velocity curve from individual component of the MW. In Section 2, we used the MWPotential2014, a combination of three gravitational potentials, to initially explore the escape velocity curve of the MW. To account for a more realistic scenario that the stellar part can be decomposed into two subcomponents, a thin and a thick disk, we assumed model I of E. Pouliasis et al. (2017), which is consistent with recent studies (e.g., A. J. Deason et al. 2019; L. Necib & T. Lin 2022a; C. Roche et al. 2024). In this model, the MW potential includes four major components, i.e., the bulge, thin and thick disks, and the DM halo. The details are described as follows.

(1) For the bulge, we use the spherical Plummer potential (H. C. Plummer 1911):

$$\Phi(r_{GC}) = -\frac{GM_{bulge}}{\sqrt{r_{GC}^2 + b^2}}, \quad (9)$$

where $M_{bulge} = 1.067 \times 10^{10} M_\odot$ and $b = 0.3$ kpc are the bulge mass and the scale constant, respectively, which is set from model I of E. Pouliasis et al. (2017).

(2) Both the thin and thick disks are modeled as the Miyamoto–Nagai profile (M. Miyamoto & R. Nagai 1975):

$$\Phi(R, z) = -\frac{GM_{disk}}{\sqrt{R^2 + (a_{disk} + \sqrt{z^2 + b_{disk}^2})^2}}, \quad (10)$$

where $R$ is the radius in the Galactocentric cylindrical coordinates, $M_{disk}$ is the mass of the disk, and $a_{disk}$ and $b_{disk}$ are the scale length and scale height, respectively. For the thin disk, we assume a mass of $M_{thin\,disk} = 3.944 \times 10^{10} M_\odot$, $a_{thin\,disk} = 5.3$ kpc, and $b_{thin\,disk} = 0.25$ kpc. For the thick disk, we adopt a mass of $M_{thick\,disk} = 3.944 \times 10^{10} M_\odot$, $a_{thick\,disk} = 2.6$ kpc, and $b_{thick\,disk} = 0.8$ kpc.

(3) The potential of the DM halo is represented by a NFW profile (J. F. Navarro et al. 1996, 1997), in which the potential of the halo is given by:

$$\Phi_{NFW}(r_{GC}) = -\frac{4\pi G \rho_0 r_{200}^3}{c^3 r_{GC}} \ln\left(1 + \frac{c r_{GC}}{r_{200}}\right), \quad (11)$$

where $c = r_{200}/r_s$ is the concentration parameter, $\rho_0 = \frac{200 \rho_{crit} \Omega_m}{3} \frac{c^3}{\ln(1+c) - c/(1+c)}$ is the characteristic density, $\Omega_m = 1.0$ is the contribution of matter to the critical density, and $r_s$ is the scale radius. The virial mass $M_{200}$ can be expressed as:

$$M_{200} = \frac{4\pi}{3} 200 \rho_{crit} \Omega_m r_{200}^3. \quad (12)$$

Therefore, the NFW profile can be described using only two parameters, $M_{200}$ and $c_{200}$.

### 5.2. Mass of the Milky Way Halo

We then constrain the mass of the MW halo with the posterior of $v_{esc}$. During fitting, we fixed the contribution of the baryonic components, i.e., the parameters of the bulge, thin disk, and thick disk, and only varied the potential of the DM halo described by $M_{200}$ and $c_{200}$. To fully explore the range of $M_{200}$ and $c_{200}$, we again applied the Bayesian theorem, leading to the posterior as follows:

$$P(M_{200}, c_{200}|v_{obs}) \propto \prod_i^{N=13} P(M_{200}) P(c_{200}) P(v_{obs,i}|M_{200}, c_{200}), \quad (13)$$

where $P(M_{200})$ and $P(c_{200})$ represent the priors on $M_{200}$ and $c_{200}$, and $P(v_{obs,i}|M_{200}, c_{200})$ stands for the likelihood of an individual observation. We assumed a uniform ($\mathcal{U}$) prior in $\log_{10}(M_{200})$ as $\mathcal{U}(11.0 < \log_{10}(M_{200}) < 12.5)$, and a uniform prior in $c_{200}$ as $\mathcal{U}(1.0 < c_{200} < 30.0)$. We also obtained the likelihood by converting the likelihood of $v_{obs,i}$ into a likelihood in the $M_{200} - c_{200}$ parameter space. For each given $M_{200}$ and $c_{200}$, we calculated the theoretical escape velocity or the circular velocity, obtained by modeling the MW potential as a model in Section 5.1. The modeling was performed using `galpy` (J. Bovy 2015).

To better constrain the parameters for the NFW halo, we also combined the escape velocities with the local circular velocity, which is consistent with previous studies (T. Piffl et al. 2014; G. Monari et al. 2018; H. H. Koppelman & A. Helmi 2021; L. Necib & T. Lin 2022a; Z. Prudil et al. 2022; C. Roche et al. 2024). In this case, $v_{obs,i}$ includes the calculated escape velocity at the central radius of each radial bin and the circular velocity at the solar position. Since the measurements of Bin 1 show trends similar to those of Bin 2, we only used the measurements of Bin 1 for the fits and adopted the full posterior distributions of $v_{esc}$ as the likelihood. For the local circular velocity, we adopted the recent measurement of Y. Zhou et al. (2023), namely $v_{circ}(R_\odot) = 234.04 \pm 10$ km s$^{-1}$, and assumed a regular likelihood function (see more details in R. J. Barlow 2019). The fits were also performed in `emcee` (D. Foreman-Mackey et al. 2013) to calculate the maximum of the log-likelihood. We used 100 walkers with 2000 steps as a burn-in and ran the walkers for additional 3000 steps.

The best-fit results are shown in Figure 4. We present the posterior distributions for the parameters $\log_{10}(M_{200})$ and $c_{200}$ in the left panel, and the escape velocity curve predicted by the best-fit model along with the measured escape velocity curve in the right panel. The final estimates for the NFW halo parameters are $\log_{10}(M_{200}[M_\odot]) = 11.91^{+0.04}_{-0.04}$ ($M_{200} = 0.81^{+0.06}_{-0.07} \times 10^{12} M_\odot$) and $c_{200} = 13.47^{+1.85}_{-1.70}$, corresponding to a total MW mass of $M_{200,\,total} = 0.90^{+0.06}_{-0.07} \times 10^{12} M_\odot$, and $r_{200} = 192.29^{+4.63}_{-5.71}$ kpc. Additional testing assuming different baryonic models shows that the choice of baryonic model has a relatively limited influence on the final results. This may be attributable to the fact that these models are all constrained by observations of the MW. We note that the uncertainty found in this study is relatively small, which implies that an escape velocity extending to the outer region can help us better constrain the DM halo mass distribution (see also, Z. Prudil et al. 2022).

According to previous simulations (e.g., R. J. J. Grand et al. 2019; H. H. Koppelman & A. Helmi 2021), the calculated escape velocities may be underestimated by 7%–10% due to the contribution of the substructures. A recent series of works have taken into account the contribution of substructures to the velocity distribution and treated it as an additional power-law component (L. Necib & T. Lin 2022a, 2022b; C. Roche et al. 2024). Given the limited sample size at larger radii and the sufficiency of a single power-law model, we did not attempt to eliminate the substructures in our high-velocity K-giant sample. To account for the influence of the substructures at larger radii more robustly, a larger sample with full 6D information reaching larger radii is necessary and we hope that such a sample will be available with the future release data of large survey projects. Nevertheless, we also obtained a corrected mass estimate by artificially increasing our escape velocities by 10%, and the NFW parameter estimates will become $M_{200,\,total} = 1.19^{+0.05}_{-0.05} \times 10^{12} M_\odot$ and $c_{200} = 13.74^{+1.30}_{-1.23}$.

### 5.3. Comparison with Previous Studies

The combination of escape velocity and local circular velocity measurements is one of the most direct methods to constrain the total mass of the MW. With this method,

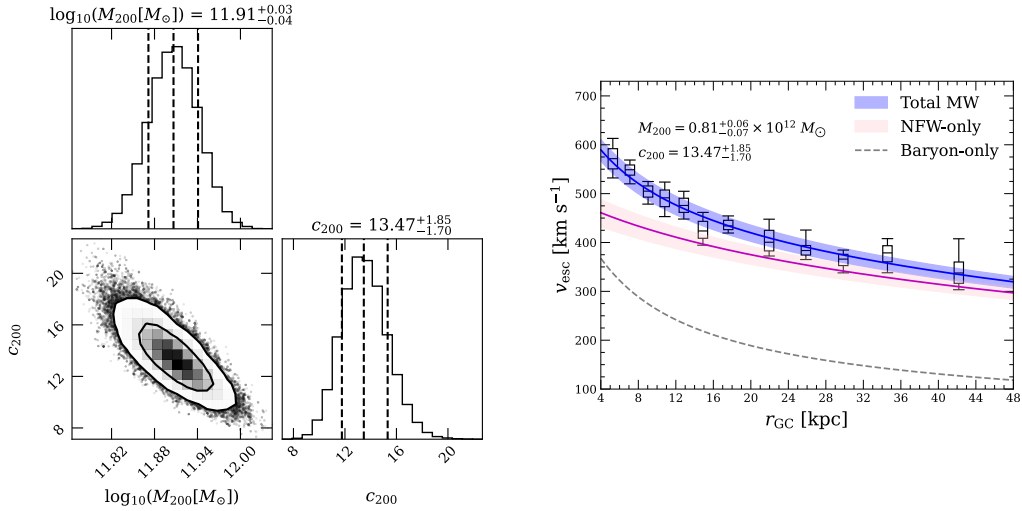

**Figure 4.** Left panel: Posterior distributions for the fitted parameters $\log_{10}(M_{200})$ and $c_{200}$ (combined the escape velocity and the local circular velocity). The 68% and 95% credible intervals are shown as contours. The numbers above the histograms represent the median parameter values and uncertainties corresponding to the 68% confidence levels. Right panel: Best fit of the total mass model to the escape velocity curve, assuming a fixed baryonic potential and NFW DM halo in the MW. The measurements are shown at 68% (boxes) and 95% (whiskers) confidence levels. The NFW parameters $M_{200}$ and $c_{200}$ are shown with 68% confidence intervals. The baryon-only and NFW-only curves show the escape velocity curves predicted by the model containing only the baryon component and DM contribution, respectively.

M. C. Smith et al. (2007) and T. Piffl et al. (2014) have provided some of the largest estimates in the past ($M_{200,\text{total}} \gtrsim 1.5 \times 10^{12} M_\odot$). However, recent mass estimates tend to be close to or below $10^{12} M_\odot$ (e.g., A. J. Deason et al. 2019; R. J. J. Grand et al. 2019; H. H. Koppelman & A. Helmi 2021; L. Necib & T. Lin 2022a; Z. Prudil et al. 2022; C. Roche et al. 2024). Table 3 lists recent total mass estimates using Gaia eDR3/DR3 with various methods. Compared with the mass estimates through the escape velocity, our results show agreement with Z. Prudil et al. (2022), which is based on the escape velocity of RR Lyrae variables, and consistent with L. Necib & T. Lin (2022a) and C. Roche et al. (2024) within the $3\sigma$ range, which have introduced a new formalism to determine the escape velocity.

Apart from escape velocity, the mass of the MW has been estimated by numerous methods, obtaining the MW mass between $0.5 \times 10^{12} M_\odot$ and $2.0 \times 10^{12} M_\odot$ (see, e.g., W. Wang et al. 2020; S. A. Bird et al. 2022; T. Sawala et al. 2023). As listed in Table 3, some mass measurements using other probes, such as measuring the circular velocity curve of the inner region (e.g., Y. Jiao et al. 2023; Y. Zhou et al. 2023; X. Ou et al. 2024; F. Sylos Labini 2024), comparing the satellite proper motions with simulations (M. K. Rodriguez Wimberly et al. 2022), modeling the kinematics of globular clusters (J. Wang et al. 2022), and using the dynamics of halo stars (A. J. Deason et al. 2021; J. Shen et al. 2022), have also tended to the mass of the MW below or close to $10^{12} M_\odot$. It is worth noting that recent measurements of the MW rotation curve reveal a gradual decline at large radii (e.g., Y. Jiao et al. 2023; H.-F. Wang et al. 2023; X. Ou et al. 2024), which implies that the MW mass should be lower than that for a flat rotation curve. To better describe the declining rotation curve, X. Ou et al. (2024) estimated the virial mass using two different DM mass models, $1.81^{+0.06}_{-0.05} \times 10^{11} M_\odot$ with an Einasto profile and $6.94^{+0.12}_{-0.11} \times 10^{11} M_\odot$ with a generalized NFW (gNFW) model. Our estimated total mass is slightly higher than that obtained by X. Ou et al. (2024) using the gNFW model. Nevertheless, our mass estimates are

**Table 3**
Comparison with Mass Estimates Using Gaia eDR3/DR3

| Method | References | $M_{200,\text{total}}$ $(10^{12} M_\odot)$ |
|---|---|---|
| Escape velocity | Our work | $0.90^{+0.06}_{-0.07}$ |
| … | … | $1.19^{+0.05}_{-0.05}$ (corrected) |
| … | C. Roche et al. (2024) | $0.64^{+0.15}_{-0.14}$ |
| … | Z. Prudil et al. (2022) | $0.83^{+0.29}_{-0.16}$ |
| … | … | $1.26^{+0.40}_{-0.22}$ (corrected) |
| … | L. Necib & T. Lin (2022a) | $0.46^{+0.11}_{-0.06}$ |
| Circular velocity curve | F. Sylos Labini (2024) | $0.65^{+0.05}_{-0.05}$ |
| … | … | $0.17^{+0.02}_{-0.02}$ (DMD) |
| … | X. Ou et al. (2024) | $0.694^{+0.012}_{-0.011}$ |
| … | … | $0.181^{+0.006}_{-0.005}$ (Einasto) |
| … | Y. Jiao et al. (2023) | $0.206^{+0.024}_{-0.013}$ |
| … | Y. Zhou et al. (2023) | $0.847 \pm 0.115$ |
| Distribution function | G. Sun et al. (2023) | $1.11^{+0.25}_{-0.18}$ |
| … | J. Wang et al. (2022) | $0.784^{+0.308}_{-0.197}$ (Zhao) |
| … | … | $0.58^{+0.081}_{-0.068}$ (Einasto) |
| … | J. Shen et al. (2022) | $1.08^{+0.12}_{-0.11}$ |
| … | A. J. Deason et al. (2021) | $1.01^{+0.24}_{-0.24}$ |
| Stream | R. Ibata et al. (2024) | $1.09^{+0.19}_{-0.14}$ |
| … | S. E. Koposov et al. (2023) | $0.774^{+0.414}_{-0.148}$ |
| Satellite proper motion | M. K. Rodriguez Wimberly et al. (2022) | $1.1^{+0.1}_{-0.1}$ |

**Notes.**
[a] The mass estimates are mostly provided assuming the NFW or gNFW profiles (J. F. Navarro et al. 1996, 1997), unless the DM model is otherwise marked. "DMD," "Einasto," and "Zhao" correspond to the DM disk model, Einasto profiles (J. Einasto 1965) and Zhao profiles (H. Zhao 1996), respectively.
[b] The corrected masses in our work and Z. Prudil et al. (2022) are both calculated by taking an increase of approximately 10% in $v_{\text{esc}}$ for the likely bias of substructures.

consistent with recent studies using the NFW or gNFW model, implying that the mass of the DM halo may be lower than in the past.

## 6. Conclusions

In this paper, we have measured the escape velocity curve using ~5000 high-velocity K giants selected from LAMOST DR8 and cross-matched with Gaia DR3. These K giants have full 6D phase-space information and relatively good quality. To expand the sample of high-velocity stars at larger radii, we adopted radius-dependent criteria, i.e., $v_{GC} > 300$ km s$^{-1}$ for the solar neighborhood ($r_{GC} \in [4, 12]$ kpc); $v_{GC} > v_{min} \sim 0.6 \times v_{esc}(r_{GC})$ for outer region ($r_{GC} > 12$ kpc). We also selected halo stars with the criteria of metallicity and $v_\phi$, which is suitable for the method of P. J. T. Leonard & S. Tremaine (1990). With this method, the tail of the velocity distribution is modeled as a power-law function to determine $v_{esc}$ at each radial bin.

For the first time, we have obtained a relatively continuous escape velocity curve extending to ~50 kpc. Viewed in the overall escape velocity curve, the escape velocity drops sharply in the inner region ($r_{GC} \lesssim 20$ kpc) and shows a smooth decline at the outer region ($r_{GC} \gtrsim 20$ kpc). Our measurements yield $v_{esc}(r_\odot) = 523.74^{+12.83}_{-13.47}$ km s$^{-1}$, which is lower than some of the previous studies (see, e.g., T. Piffl et al. 2014; G. Monari et al. 2018), but is well consistent with some of the most recent ones (see, e.g., H. H. Koppelman & A. Helmi 2021; L. Necib & T. Lin 2022a; Z. Prudil et al. 2022; C. Roche et al. 2024). For the outer region, our results are similar to A. A. Williams et al. (2017) and Z. Prudil et al. (2022).

Based on the resulting escape velocity curve and the recent circular velocity measurements (Y. Zhou et al. 2023), we constrained the DM halo parameters by assuming a mass model that included the baryonic components and the DM halo described by the NFW model. The final estimates are a DM halo mass of $M_{200} = 0.81^{+0.06}_{-0.07} \times 10^{12} M_\odot$, with a concentration of $c_{200} = 13.47^{+1.85}_{-1.70}$, corresponding to a total MW mass of $M_{200, \text{total}} = 0.90^{+0.06}_{-0.07} \times 10^{12} M_\odot$, and $r_{200} = 192.29^{+4.63}_{-5.71}$ kpc. The relatively small uncertainty implies that it is valuable to include escape velocity measurements from the outer region. Compared with more previous mass estimates using the solar escape velocity (M. C. Smith et al. 2007; T. Piffl et al. 2014; G. Monari et al. 2018), our derived MW mass may support a lower mass of the DM halo, which is consistent with some recent studies using other tracers (e.g., W. Wang et al. 2020; S. A. Bird et al. 2022; T. Sawala et al. 2023, for recent reviews).

## Acknowledgments

We thank the anonymous referee for the very efficient and helpful report to improve the draft. This work is supported by the CAS Project for Young Scientists in Basic Research (Nos. YSBR-092 and YSBR-062), the Strategic Priority Research Program of Chinese Academy of Sciences grant Nos. XDB1160100, XDB1160102 and XDB34020205, the National Natural Science Foundation of China (NSFC) grant Nos. 12222305, 12422303 and 12588202, the National Key R&D Program of China Nos. 2024YFA1611902, 2024YFA1611903 and 2023YFE0107800, and the science research grants from the China Manned Space Project with No. CMS-CSST-2025-A11.

Guoshoujing Telescope (the Large Sky Area Multi-Object Fiber Spectroscopic Telescope, LAMOST) is a National Major Scientific Project built by the Chinese Academy of Sciences. Funding for the project has been provided by the National Development and Reform Commission. It is operated and managed by the National Astronomical Observatories, Chinese Academy of Sciences.

This work has made use of data from the European Space Agency (ESA) mission Gaia (https://www.cosmos.esa.int/gaia), processed by the Gaia Data Processing and Analysis Consortium (DPAC, https://www.cosmos.esa.int/web/gaia/dpac/consortium). Funding for the DPAC has been provided by national institutions, in particular the institutions participating in the Gaia Multilateral Agreement.

*Software:* Numpy (C. R. Harris et al. 2020), scipy (P. Virtanen et al. 2020), astropy (Astropy Collaboration et al. 2013, 2018, 2022), corner.py (D. Foreman-Mackey 2016), emcee (D. Foreman-Mackey et al. 2013), galpy (J. Bovy 2015).

## Appendix A
## Additional Tables

Table A1 lists the measured $v_{esc}$ in all redial bins for Bin 1 and Bin 2. The radial bins are marked as the median value of the observed Galactocentric radius for the stars in each radial bin, e.g., 5.29 kpc for 4–6 kpc radial bin. The errors are 68% credible intervals calculated by MCMC.

**Table A1**
Escape Velocity Measurements in Different Bins

| Bin 1 | | | Bin 2 | | |
|---|---|---|---|---|---|
| $r_{GC}$ (kpc) | $v_{esc}$ (km s$^{-1}$) | $N_{stars}$ | $r_{GC}$ (kpc) | $v_{esc}$ (km s$^{-1}$) | $N_{stars}$ |
| 5.29 | $571.57^{+20.42}_{-20.89}$ | 167 | 5.70 | $564.80^{+15.21}_{-16.13}$ | 258 |
| 7.09 | $548.93^{+9.68}_{-11.09}$ | 420 | 7.55 | $549.70^{+9.64}_{-10.45}$ | 437 |
| 9.02 | $505.06^{+10.67}_{-12.20}$ | 506 | 9.36 | $491.20^{+12.54}_{-13.31}$ | 510 |
| 10.83 | $491.44^{+15.94}_{-18.04}$ | 335 | 11.34 | $493.37^{+13.60}_{-14.59}$ | 280 |
| 12.84 | $475.77^{+14.78}_{-14.14}$ | 320 | 13.38 | $472.28^{+14.18}_{-15.73}$ | 280 |
| 14.90 | $423.59^{+19.68}_{-14.99}$ | 212 | 15.32 | $412.90^{+21.39}_{-13.38}$ | 178 |
| 17.59 | $433.96^{+11.44}_{-7.65}$ | 306 | 18.09 | $427.52^{+8.07}_{-7.92}$ | 274 |
| 21.97 | $400.49^{+24.39}_{-17.73}$ | 191 | 22.40 | $401.45^{+24.06}_{-18.32}$ | 179 |
| 25.88 | $383.59^{+10.20}_{-9.29}$ | 168 | 26.42 | $384.00^{+9.44}_{-8.68}$ | 167 |
| 29.86 | $365.99^{+9.75}_{-13.23}$ | 141 | 30.20 | $374.69^{+7.28}_{-6.17}$ | 132 |
| 34.57 | $378.72^{+14.69}_{-18.32}$ | 156 | 34.91 | $370.46^{+18.71}_{-21.44}$ | 144 |
| 42.06 | $337.59^{+22.46}_{-21.15}$ | 55 | 42.27 | $342.88^{+48.91}_{-22.38}$ | 52 |

## Appendix B
## Additional Figures

The figures in this appendix are organized as follows. Figures B1 and B2 are best-fit results for all radial bins of Bin 1 and Bin 2, respectively. We only presented the favored results after the comparison of the AIC. Figures B3 and B4 show the posterior distributions of the fitted parameters for representative radial bins at 8–10 kpc and 14–16 kpc, respectively.

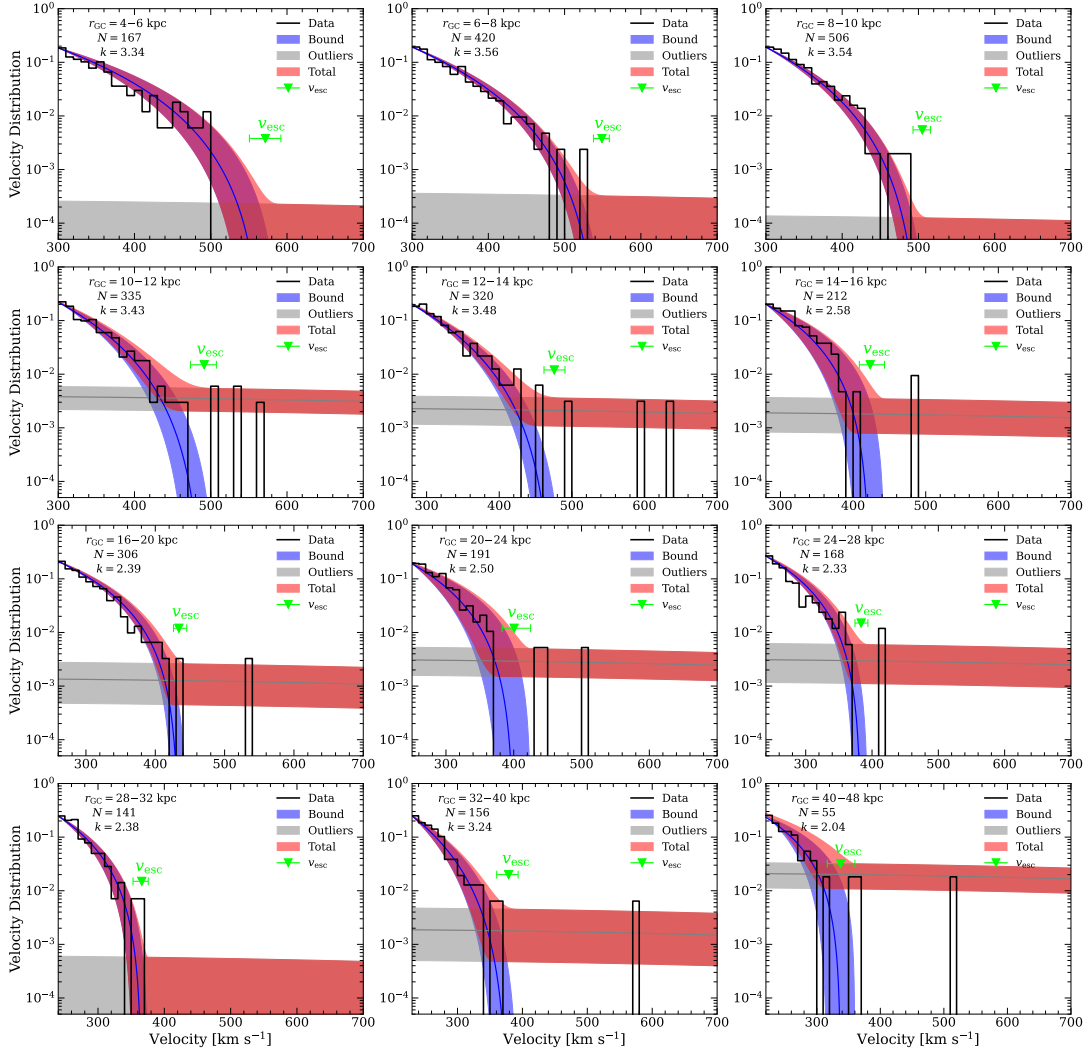

**Figure B1.** Best-fit results for all radial bins of Bin 1 using $v_{\min}$ in Table 1. The black solid line is the velocity distribution of stars, while the shaded regions are the distribution predicted by the model with 68% confidence intervals. The contributions of the bound and outlier components are shown individually. The $v_{\rm esc}$ marker with error bars represents the best-fit escape velocity and its 68% confidence intervals.

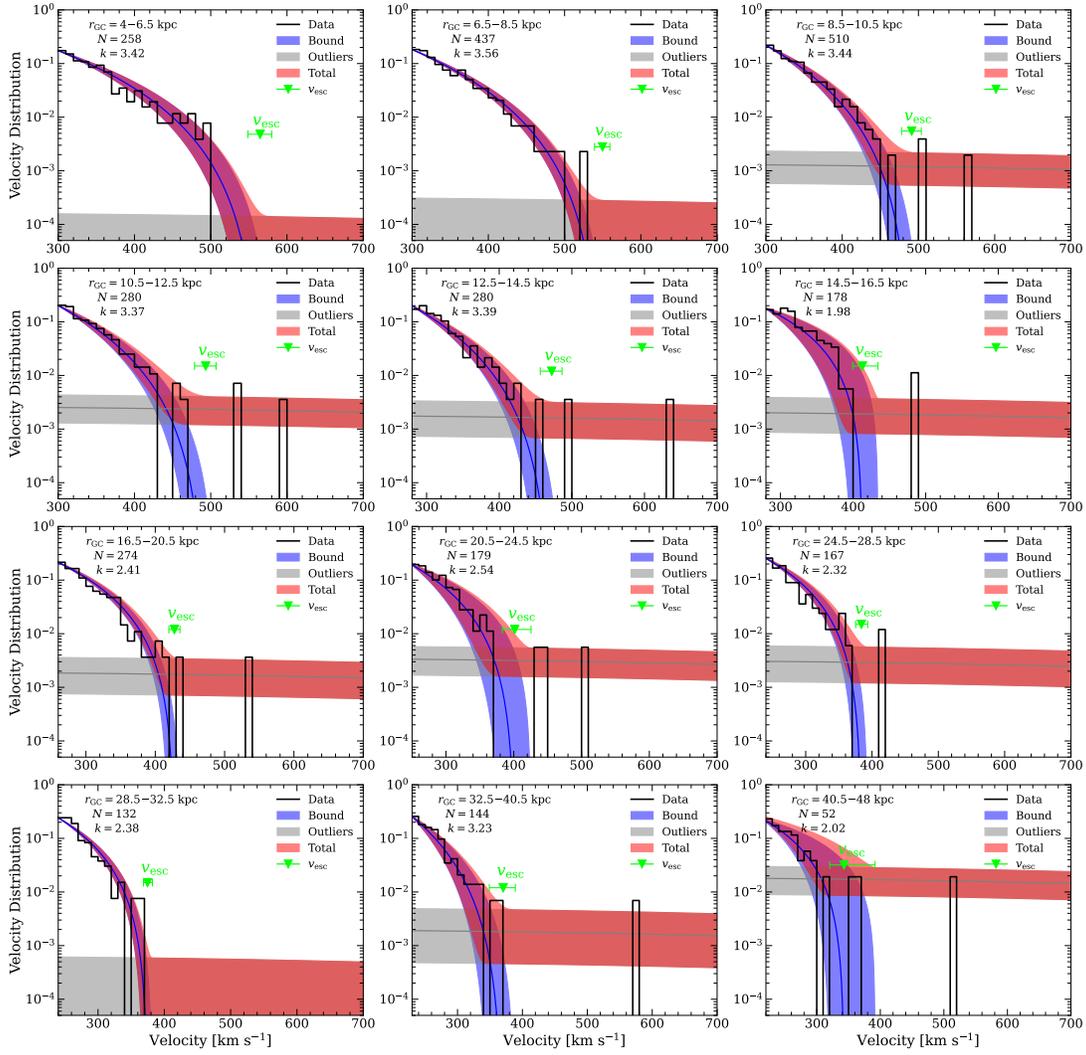

**Figure B2.** Best-fit results for all radial bins of Bin 2 using $v_{\rm min}$ in Table 1. The black solid line is the velocity distribution of stars, while the shaded regions are the distribution predicted by the model with 68% confidence intervals. The contributions of the bound and outlier components are shown individually. The $v_{\rm esc}$ marker with error bars represents the best-fit escape velocity and its 68% confidence intervals.

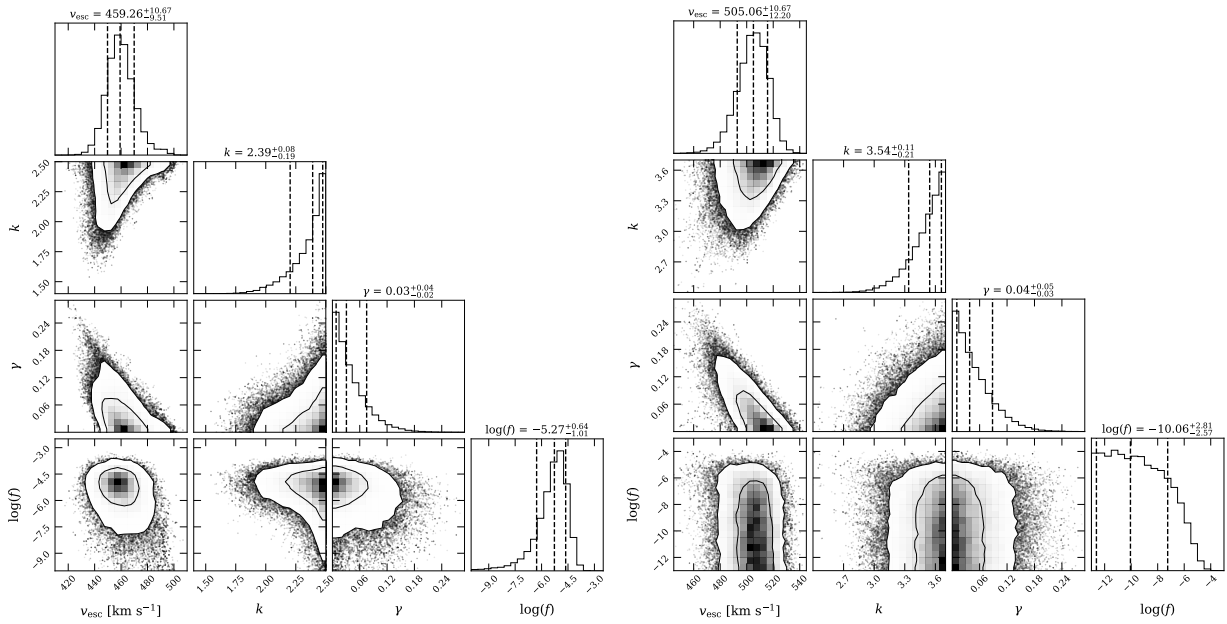

**Figure B3.** Posterior distributions of the fitted parameters $v_{esc}$, $k$, $\gamma$ and $f$ for the 8–10 kpc radial bin, which adopted $v_{min} = 300$ km s$^{-1}$. The left and right panels correspond to $k \in [0.1, 2.5]$ and $k \in [0.1, 3.7]$, respectively. The 68% and 95% credible intervals are shown as contours. The numbers above the histograms represent the median parameter values and uncertainties corresponding to the 68% confidence levels.

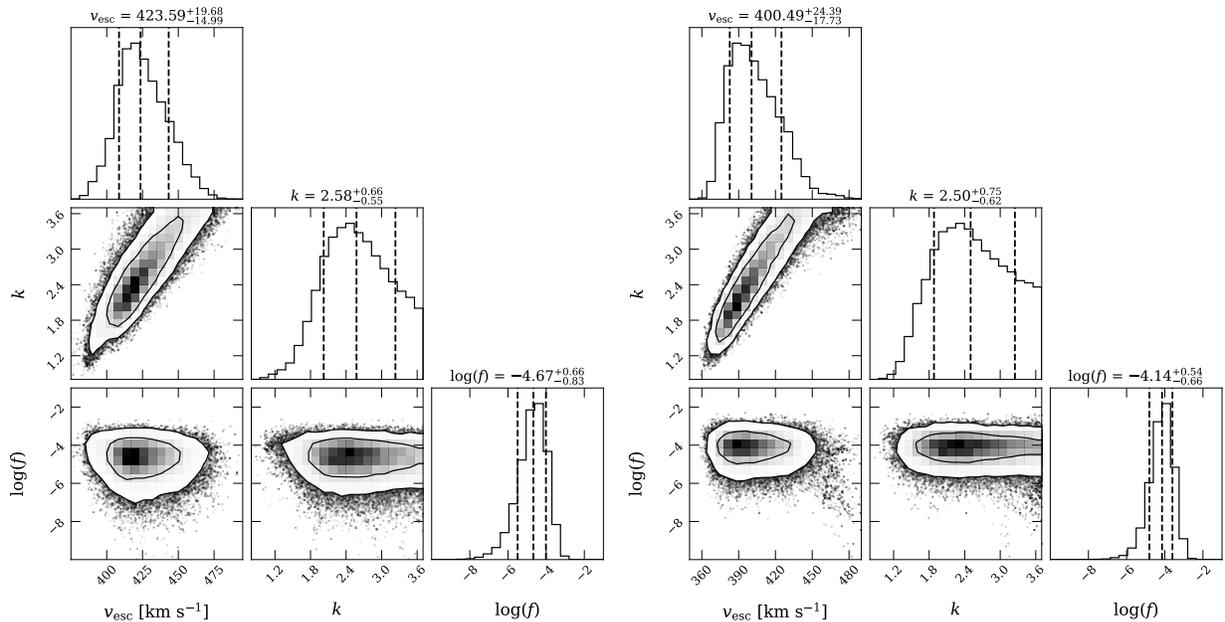

**Figure B4.** Posterior distributions of the fitted parameters $v_{esc}$, $k$ and $f$ when adopting $k \in [0.1, 3.7]$ for the 14–16 kpc radial bin (left panel, $v_{min} = 280$ km s$^{-1}$) and the 20–24 kpc radial bin (right panel, $v_{min} = 250$ km s$^{-1}$). The 68% and 95% credible intervals are shown as contours. The numbers above the histograms represent the median parameter values and uncertainties corresponding to the 68% confidence levels.

## ORCID iDs

Yin Wu ● https://orcid.org/0009-0003-4116-6824
Haining Li ● https://orcid.org/0000-0002-0389-9264
Yang Huang ● https://orcid.org/0000-0003-3250-2876
Xiang-Xiang Xue ● https://orcid.org/0000-0002-0642-5689
Gang Zhao ● https://orcid.org/0000-0002-8980-945X